\def\one{|1\rangle}

\def\zero{|0\rangle}

\def\gr{|g\rangle}

\def\ex{|e\rangle}

\def\tx{\tilde\sigma_x}

\def\tz{\tilde\sigma_z}
\def\sx{\sigma_x}
\def\sy{\sigma_y}
\def\sz{\sigma_z}

\def\x1{\chi_z}
\def\x2{\chi_{zz}}

\def\w{\omega}
\def\w0{\omega_0}
\def\W{\Omega}
\def\W0{\Omega_0}

\documentclass[prl,twocolumn,superscriptaddress]{revtex4}
\usepackage{graphicx}
\usepackage[dvips]{color}
\begin{document}
\title{Quantum two-level systems in Josephson junctions as naturally formed
qubits}

\author{A. M. Zagoskin}
\affiliation{Frontier Research System, The Institute of Physical
and Chemical Research (RIKEN), Wako-shi, Saitama, Japan}
\affiliation{Department of Physics and Astronomy, The University
of British Columbia, Vancouver, B.C., Canada}
\author{S. Ashhab}
\affiliation{Frontier Research System, The Institute of Physical
and Chemical Research (RIKEN), Wako-shi, Saitama, Japan}
\author{J. R. Johansson}
\affiliation{Frontier Research System, The Institute of Physical
and Chemical Research (RIKEN), Wako-shi, Saitama, Japan}
\author{Franco Nori}
\affiliation{Frontier Research System, The Institute of Physical
and Chemical Research (RIKEN), Wako-shi, Saitama, Japan}
\affiliation{MCTP, CSCS, Department of Physics, The University of
Michigan, Ann Arbor, Michigan, USA}

\begin{abstract}
The two-level systems (TLSs) naturally occurring in Josephson
junctions constitute a major obstacle for the operation of
superconducting phase qubits. Since these TLSs can possess
remarkably long decoherence times, we show that such TLSs can
themselves be used as qubits, allowing for a well controlled
initialization, universal sets of quantum gates, and readout.
Thus, a single current-biased Josephson junction (CBJJ) can be
considered as a multiqubit register. It can be coupled to other
CBJJs to allow the application of quantum gates to an arbitrary
pair of qubits in the system. Our results indicate an alternative
way to realize superconducting quantum information processing.
\end{abstract}

\maketitle

Several advances in the field of quantum information processing
using superconducting circuits have been made in recent years
\cite{You}. A major obstacle to further advances, however, is the
problem of decoherence. In particular, recent experiments on
current-biased Josephson junctions (CBJJs) revealed resonances
that suggest the presence of quantum two-level systems (TLSs) that
are strongly coupled to the CBJJ when it is biased near one of
those resonances \cite{Simmonds,MS+S}. So far the TLSs have been
treated as a nuisance that prevent the operation of the CBJJ as a
qubit near any resonance. In this paper we show that {\em the TLSs
themselves can be used as qubits}. The CBJJ then acts as a bus,
enabling state initialization, one- and two-qubit operations, and
readout. Moreover, the results of Refs.~\cite{Simmonds,MS+S} show
that the decoherence times of the TLSs are longer than those of
the CBJJ. That property can be used in a scalable design such that
the decoherence time of the entire system scales as the CBJJ
decoherence time $T_{\rm d}^{\rm CBJJ}$, {\it as opposed to the
usual} $T_{\rm d}^{\rm CBJJ}/N$, where $N$ is the number of CBJJs
in the circuit. Our results therefore demonstrate an alternative
way to achieve a scalable qubit network in a superconducting
system, in addition to illustrating a method to perform
multi-qubit experiments with available experimental capabilities.

{\it Model.---\/}The phase qubit, which is comprised of a single
CBJJ, is one of the simplest experimental implementations of a
qubit in superconducting systems \cite{You,Martinis_0}. The
working states, $\zero$ and $\one$, are the (metastable) ground
and first excited states in a local minimum of the washboard
potential produced by the CBJJ. The (undriven) system is described
by the Hamiltonian
\begin{equation}
H = \frac{\hat{Q}^2}{2C} - \frac{I_c\Phi_0}{2\pi}\cos\hat{\varphi}
- \frac{I_b\Phi_0}{2\pi}\hat{\varphi} \approx
\frac{\hbar\omega_{10}}{2}\sz. \label{eq_1}
\end{equation}
Here $\hat{Q}$ is the operator of the charge on the junction, $C$
is the junction's capacitance, $\hat{\varphi}$ is the operator of
the Josephson phase difference, and $I_c, I_b (\leq I_c)$ are the
critical and bias current, respectively. The nonlinearity of the
potential allows to consider only the two lowest energy  states.
The Pauli matrix $\sz$ operates in the subspace $\{\zero,\one\}$.
The transition frequency $\omega_{10} \approx \omega_p = (2\pi
I_c/\Phi_0 C)^{1/2} (1-j^2)^{1/4}$, the plasma frequency in the
biased junction (the corrections due to nonlinearity are $\sim
10\%$ \cite{Martinis_1}), and $j=I_b/I_c$ (see
e.g.~\cite{Martinis_0}). The terms in the Hamiltonian proportional
to $\sx$ and $\sy$, which enable a complete set of one-qubit
gates, appear in the rotated frame of reference when applying
microwave pulses of bias current at the resonance frequency, $I_b
\to I_{\rm DC}+ I_{\mu wc}(t) \cos(\omega_{10}t)+I_{\mu ws}(t)
\sin(\omega_{10}t)$, as explained in Ref.~\cite{Martinis_1}.

The  simplicity of the qubit design and manipulation contributed
to its successful experimental realization
\cite{You,Martinis_1,Simmonds} and a spectroscopic demonstration
of the formation of entangled two-qubit states \cite{Berkley}.
Nevertheless the experiment in Ref.~\cite{Simmonds} also
demonstrated that TLSs present in the tunneling barrier tend to
destroy the coherent operation of the qubit (such as Rabi
oscillations).

Such TLSs are ubiquitous in solid state systems wherever disorder
is present and can be thought of as  groups of atoms capable of
tunneling through a potential barrier between two degenerate
configurations. They are currently believed to be the main source
of $1/f$-noise in solids.

The observed  coherent oscillations between a TLS and a phase
qubit \cite{Cooper} proved that a TLS can be considered as a
coherent quantum object described by the pseudospin Hamiltonian
\begin{equation}
H_{\rm TLS} = - \; \frac{\Delta}{2}\tx - \, \frac{\epsilon}{2}\tz,
\label{eq_TLS}
\end{equation}
with a decoherence time  {\em longer} than that of the qubit. The
Pauli matrices $\tx$ and $\tz$ operate on the TLS states. Note
that, since the nature of the TLSs is currently unknown, one
cannot derive the values of the TLS parameters from first
principles. As will become clear below, however, neither a
derivation of those parameters nor an understanding of their
physical origin is necessary in order to make use of the TLSs.

Although the available experimental
data gives rather limited information  
about the TLS-TLS interaction, it is highly unlikely that the TLSs
interact directly with each other or with the external fields,
because of their supposedly microscopic dipole moments and
relatively large spatial separation.

The TLS-CBJJ coupling is believed to be due to one of the
following mechanisms: (A) through the critical current dependence
on the TLS position \cite{Simmonds} or (B) the direct dipole
coupling to the junction charge $\hat{Q}$ (which is currently
considered more likely) \cite{MBS,MS+S}:
\begin{eqnarray}
\hat{H}_{\rm int}^{(A)} &=& - \, \frac{I_c \Phi_0}{2\pi}
\frac{\delta I_c}{2I_c} \cos \phi \, \tz;\label{eq:Hint}\\
\hat{H}_{\rm int}^{(B)} &=& \alpha \, \hat{Q} \, \tz . \nonumber
\end{eqnarray}
Both produce the coupling term $h_x \sx\tz$. In addition, in the
former case there appears an $h_z\sz\tz$-term, which reflects the
change in the inter-level spacing  $\omega_{10}$ due to the plasma
frequency dependence on the bias. The ratio $\gamma \equiv h_z/h_x
= (E_C/2E_J)^{1/4}(1-j^2)^{-5/8}$, where $E_C=2e^2/C$ and $E_J =
I_c\Phi_0/2\pi$.  If $(E_C/2E_J)=10^{-7}$ (achievable using the
external capacitor technique \cite{MS+S}) and $j=0.90$, then
$\gamma=0.05$ \cite{Temperature}.

Hereafter, we consider the case where $\gamma  \ll 1$. Otherwise,
the effective TLS-CBJJ decoupling is impossible even when out of
resonance (on the positive side, this effect would provide the
means to definitively establish the mechanism of TLS-CBJJ
coupling). Then, performing a basis transformation on the TLS
states, relabeling $[\Delta^2+\epsilon^2]^{1/2}\to\Delta$, and
using the rotating-wave approximation, we finally obtain the
effective Hamiltonian for the TLS-CBJJ  system:
\begin{equation}
\hat{H} = -\frac{\hbar\omega_{10}}{2} \sz - \sum_j \left(
\frac{\Delta_j}{2} \tz^j + \lambda_j \sx \tx^j \right),
\label{eq_4}
\end{equation}
\noindent with the effective coupling coefficients $\lambda_j$.
The coupling term acts only in resonance, when $|\hbar\omega_{10}
- \Delta_j| < \lambda_j$.

The Hamiltonian (\ref{eq_4}) was derived under the experimentally
relevant assumption that the coupling $\lambda \ll
\hbar\omega_{10} \sim \Delta$. We neglected  a term of the form
$\mu\sx\tz^j$ in Eq.~(\ref{eq_4}), because its influence on the
system's dynamics is negligible when $\mu \ll \hbar\omega_{10}$
and $\mu^2/\hbar\omega_{10} \ll \lambda$.

{\it Qubit operations.---\/} The inter-level spacing
$\hbar\omega_{10}$ is tuned by the bias current, which together
with the resonant behaviour of the TLS-CBJJ coupling allows the
independent manipulation of each TLS (due to the natural
dispersion of their characteristic energies, $\Delta_j$). The
condition $\lambda\ll\Delta,\hbar\omega_{10}$ allows us to
consider changes in the bias current as adiabatic from the point
of view of internal CBJJ and TLS evolution, but instantaneous with
respect to the CBJJ-TLS dynamics.

\begin{figure}[h]
\includegraphics
[width=9cm]{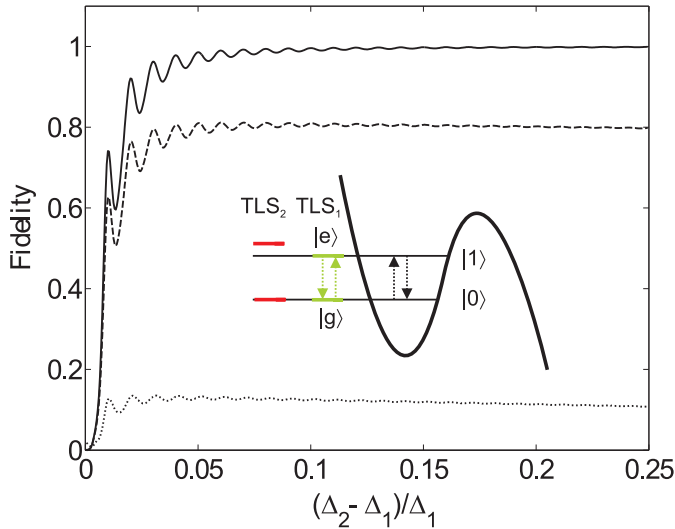} \caption{ Fidelity of the ISWAP
gate on two TLS qubits in a CBJJ as a function of the relative
difference between the TLS energy splittings ($\Delta_1$ and
$\Delta_2$, respectively). The solid line corresponds to no
decoherence, the dashed and dotted lines correspond to CBJJ
decoherence rates $\Gamma_1=\lambda/20\pi$ and
$\Gamma_1=\lambda/2\pi$, respectively, where in both cases
$\Gamma_2=2\Gamma_1$. The much slower decoherence of the TLSs have
been neglected. $\lambda = 0.005\Delta_1$. (Inset) Schematic
depiction of the quantum beats  between the CBJJ and the TLS$_1$
in resonance; TLS$_2$ is effectively decoupled from the
CBJJ.}\label{fig1}
\end{figure}

First, consider the single-qubit operations, assuming for the time
being that the decoherence times of the TLSs, $T_{{\rm d}}^{(j)}$,
exceed  the decoherence time of the CBJJ, $T_{\rm d}^{\rm CBJJ}$,
which in turn is much larger than the characteristic interaction
time $\hbar/\lambda$. On resonance with the $j$th TLS, the
Hamiltonian (\ref{eq_4}) contains a block that acts as
$-\lambda_j\sx$ in the subspace of degenerate states
$\{\one\otimes\gr,\zero\otimes\ex\}.$ Its operation leads to
quantum beats between the states of the TLS and the CBJJ (see
Fig.~\ref{fig1} (inset)), with the period $\tau_j \equiv
\hbar/\lambda_j,$ as was observed in  \cite{Simmonds}. Therefore
single-qubit operations on a TLS and its initialization to an
arbitrary state can be achieved as follows: After initializing the
CBJJ in the state $\zero$ we bring it in resonance with the TLS
for the duration $\tau_j/2$; as a result the states of TLS and
qubit are swapped,
\begin{equation}
\zero\otimes(\alpha\gr+\beta\ex) \to
e^{i\frac{\pi}{2}}\left(\alpha
e^{i\frac{\pi}{2}(1+\Delta/\lambda)}\zero +
\beta\one\right)\otimes\gr. \label{eq_5}
\end{equation}
Then the CBJJ is taken out of resonance with the TLS, a rotation
of its state is performed, and the resulting state is again
transferred to the TLS (always compensating for the parasite phase
shifts $\frac{\pi}{2}(1+\Delta/\lambda)$). The readout of the
qubit state can be  performed after the swap (\ref{eq_5}), e.g. by
using the technique of \cite{Cooper}, where the transition to the
resistive state, with its potentially-problematic coupling to the
TLSs, is avoided. The decoherence time $T_{{\rm d}}^{(j)}$ can be
determined using a similar sequence of operations, by initializing
the TLS in a superposition state, decoupling it from the CBJJ, and
measuring the probability to find the TLS in a given state after a
given time. Note also that if the Rabi frequency of CBJJ greatly
exceeds the TLS-CBJJ coupling, the manipulations with the state of
the CBJJ can be performed while in resonance with a TLS, reducing
the overall operation time \cite{Ashhab}.

Note that as soon as the CBJJ is biased away from the resonance
with the TLS, they effectively decouple, and any perturbation of
the quantum state of the CBJJ does not affect the survival of
quantum coherence in the TLS. Therefore the quality factor (the
number of gates, of average duration $\tau_{\rm gate}$, that can
be performed in the system before it loses quantum coherence) is
no less than $T_{\rm d}^{\rm CBJJ}/\tau_{\rm gate}$, and can
exceed this value depending on the specific decoherence decay law.

The application of two-qubit gates between two TLSs inside the
same CBJJ can be achieved similarly. For example, the universal
ISWAP gate can be performed by the following sequence of
operations: tune the CBJJ in resonance with TLS$_1$ for
$\tau_1/2$, then tune it in resonance with TLS$_2$ for $\tau_2/2$,
and finally again in resonance with TLS$_1$ for $\tau_1/2$. (In
the process each TLS qubit was also rotated by $\pi/2$ around
$z$-axis, which can be trivially repaired by performing the
single-qubit rotation as described in the previous paragraph.) At
the end of the above sequence, the CBJJ is decoupled {\em and}
disentangled from both TLSs.

{\it Numerical estimates.---\/} The idealized picture of two-qubit
operations neglected (1) the finite detuning between TLS qubits,
(2) the influence of other TLSs present in the CBJJ, (3)
decoherence and (4) the finite time of the CBJJ bias current
adjustment. We numerically investigated  the impact of these
effects on the fidelity $F$ defined as the minimum
\begin{equation}
F= \min_{| \Psi_i \rangle}( \langle \Psi_i | \; U_{\rm
ideal}^{\dagger} \; \rho_f \; U_{\rm ideal} \; | \Psi_i \rangle )
\label{eq:Fidelity}
\end{equation}
\noindent where $| \Psi_i \rangle$ is an initial state, $U_{\rm
ideal}$ is the (ideal) desired operation, and $\rho_f$ is the
numerically obtained final density matrix. (note that the quantity
inside the parenthese in Eq. (\ref{eq:Fidelity}) depends on the
initial state).

In the numerical simulations, we took the ISWAP operation as a
representative quantum gate and used 900 different initial states.
The results are as follows (see Fig.~\ref{fig1}). For two TLS
qubits, with $\lambda_1=\lambda_2=0.005 \Delta_1$ (values close to
the experimental data \cite{Simmonds,Cooper}), the fidelity first
reaches 90\% when $\delta \equiv |\Delta_1-\Delta_2| = 3.5
\lambda$, and 99\% when $\delta \approx 10 \lambda.$ When $\delta
= 8\lambda$, the fidelity is 98.8\%. Adding an idle TLS with
$\lambda=0.002\Delta_1$ in resonance with one of the qubit TLSs
reduces the fidelity to 80\% (94\% for $\lambda=0.001\Delta_1$).
Finally, with two qubit TLSs with reduced energies 1 and 1.04,
four idle TLSs with reduced energies 0.99, 1.01, 1.03 and 1.05,
and the reduced coupling strengths $\lambda_{\rm qubit}=0.005$ and
$\lambda_{\rm idle}=0.002$, respectively, we find that the
fidelity of the ISWAP gate $\approx$ 95\%. Therefore, we
conclude that the finite detuning and presence of idle TLSs per se
is not dangerous.

We now take into account the decoherence in the CBJJ, neglecting
the much weaker one in the TLSs. The ISWAP fidelity   with
$\delta=8\lambda_j$ and without idle TLSs is 81\% when
$\Gamma_1=(1/2)\Gamma_2=\lambda/20\pi$, but it drops to 13\% when
$\Gamma_1=(1/2)\Gamma_2=\lambda/2\pi$.

Finally, we now consider the effects of a finite CBJJ bias
switching time between the  resonant frequencies of the TLSs.
Taking a simple linear  ramp with $t=2$ ns and no pulse
optimization, and neglecting decoherence, we find that the
fidelity drops to 80\%. Finite decoherence
$\Gamma_1=(1/2)\Gamma_2=\lambda/20\pi$ further suppresses it to
63\%. The above estimates show that the operation of the proposed
two-qubit gate can be realized  with the current experimental
techniques used, e.g., in Ref.~\cite{MS+S}.

\begin{figure}[b]
\includegraphics[width=9cm]{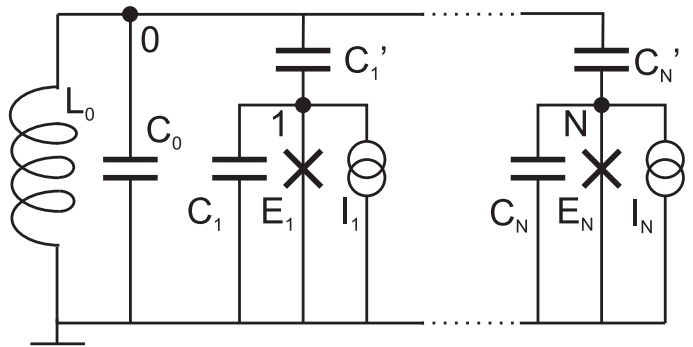}
\caption{ Scalability of the structure: Two-qubit operations
between the TLSs on different CBJJs ($j=1,\dots,N$) are enabled by
the common LC circuit, which
is capacitively coupled to the CBJJs.  The Josephson energy,  capacitance, and
bias current of the $j$th CBJJ are $E_j, C_j, I_j$ resp.
}\label{fig2}
\end{figure}

{\it Scalability.---\/} To ensure the scalability of the system,
we must be able to perform two-qubit gates  on TLSs located in
different CBJJs. It can be done by swapping the states of the
capacitively coupled adjacent CBJJs like in \cite{Blais}, but a
better solution is based on   the method suggested in
\cite{Makhlin}. Here the qubit-carrying  CBJJs are coupled
capacitively to a common linear LC circuit with   resonance
frequency $\omega_0$ much higher than the characteristic
frequencies of separate CBJJs (Fig.~\ref{fig2}).

Using the standard formalism, we can  write the Lagrangian of the
system,

\begin{eqnarray}
{\cal L} = \frac{C_0\dot{\phi}_0^2}{2} +
\sum_{j=1}^N\left[\frac{C_j\dot{\phi}_j^2}{2}
+\frac{C_j'\left(\dot{\phi}_j-\dot{\phi}_0\right)^2}{2} \right] -
\frac{\phi_0^2}{2L_0}-  \\
\sum_{j=1}^N \left[-I_j\phi_j - E_j
\cos\frac{2e\phi_j}{\hbar}\right] \equiv \frac{1}{2}
\sum_{j,k=0}^{N} {\cal C}_{jk}\dot{\phi}_j\dot{\phi}_k -
U(\{\phi\}), \nonumber
\end{eqnarray}
where $\dot{\phi}_j(t) \equiv V_j(t)$, the voltage between node
$j$ and the ground. The corresponding Hamiltonian becomes
$$
H = 
\frac{1}{2} \sum_{j,k=0}^{N} {\cal C}^{-1}_{jk}\hat{p}_j\hat{p}_k
+ \frac{1}{2} \sum_{j=0}^{N}\frac{\omega_j^2}{{\cal
C}^{-1}_{jj}}\hat{\phi_j}^2 +
\cdots,
$$
where $\hat{\phi} = (\hbar/2e)\hat{\varphi}$ (see
Eq.~(\ref{eq_1})), $[\hat{\phi}_j,\hat{p}_j]=i\hbar$, ${\cal
C}^{-1}$ is the inverse capacitance matrix, $\omega_0$ is the
frequency of the LC bus, $\omega_j$ is the frequency of the $j$th
CBJJ in the harmonic approximation, and the ellipsis stands for
the nonlinear corrections. After introducing the Bose operators
$a$ and $a^{\dag}$ via $
\hat{\phi}_j =\Lambda_j (a_j+a_j^{\dag})/2,  \hat{p}_j =
\hbar(a_j-a_j^{\dag})/(i\Lambda_j), \Lambda_j = \left((2\hbar{\cal
C}^{-1}_{jj})/\omega_j\right)^{1/2},
$ the Hamiltonian becomes
$$
H = \sum_{j=0}^N \hbar\omega_j (a_j^{\dag}a_j +\frac{1}{2}) - \!
\! \sum_{k>j=0}^N \!\!
g_{jk}\!\left(a_j-a_j^{\dag}\right)\!\!\left(a_k-a_k^{\dag}\right)
+ \cdots \label{eq_H2}
$$
with the effective coupling
$$ g_{jk} = \hbar (\omega_j\omega_k)^{1/2} {\cal C}^{-1}_{jk} /
\left[2\left( {\cal C}^{-1}_{jj} {\cal
C}^{-1}_{kk}\right)^{1/2}\right].$$

By assumption, $\omega_0 \gg \omega_j,\:j=1,\dots N$. Therefore
the Hamiltonian can be projected on the ground state of the LC bus
\cite{Makhlin}. The nonlinearity of the CBJJ allows us to further
restrict the Hamiltonian to the subspace spanned by the states
$\zero, \one$ of each CBJJ, producing
\begin{equation}
H_{\rm eff} = \frac{1}{2}\sum_{j=1}^N \hbar\omega_j \; \sz^j +
\sum_{k>j=1}^N g_{jk} \; \sy^j \; \sy^k. \label{eq_H3}
\end{equation}

In the interaction representation with respect to $H_0 =
\frac{1}{2}\sum_{j=1}^N \hbar\omega_j \sz^j$ it is easy to see
that $\sy^j(t) = \sy^j(0)\cos\omega_jt + \sx^j(0)\sin\omega_jt$
\cite{Orszag}. Therefore, the pairwise couplings in (\ref{eq_H3})
will only be effective for the CBJJ tuned in resonance with each
other, in which case ($\omega_j=\omega_k$) the effective
interaction term is $\tilde{H}_{\rm eff}^{jk} = g_{jk}
\left(\sx^j\sx^k+\sy^j\sy^k\right)/2$. On the subspace spanned by
the states $\zero_j\otimes\one_k$ and $\one_j\otimes\zero_k$ the
operator $\left(\sx^j\sx^k+\sy^j\sy^k\right)/2$ acts as $\sx$
(while it is exactly zero outside). Therefore this coupling allows
the same universal two-CBJJ manipulations as in  \cite{Blais}.
Universal two-qubit gates on TLSs situated in different CBJJs can
then be performed by transferring the states of the TLS$_{1,2}$ to
the corresponding CBJJ$_{1,2}$, performing the two-qubit
operations on the states of CBJJ$_{1,2}$, and retransferring the
resulting states back to the TLSs.

The number of TLS per CBJJ is of order 10 and depends on the
fabrication. The decoherence times of TLSs are determined by their
local environment and are insensitive to the number of CBJJs linked
to the same LC circuit. Similarly, the influence of the LC circuit
on the decoherence of the CBJJ is negligible as long as $\omega_0
\gg \omega_{10} \sim \omega_p$ \cite{Martinis_0}, the same
requirement we need to obtain the coupling $\tilde{H}_{\rm
eff}^{jk}$. Therefore the scalability of the system is limited by
the condition $[L(C_0+NC'_{1,\rm eff})]^{-1} \gg I_c/[\Phi_0 C_1],
$ or, to the same accuracy,   $N \ll
\left[\Phi_0/LI_c\right]\left[C_1/C'_1\right].$ Therefore $N$ can
be of order few dozen without violating the applicability
conditions for the above scalable design.

The usefulness of TLSs for our purpose could be questioned because
their parameters undergo spontaneous changes. Nevertheless, since
such changes typically happen on the scale of days \cite{McD}, a
pre-run calibration is sufficient for any realistic task. Another
concern that could be raised is the fact that any operation on the
TLSs is done through the CBJJ, so that the latter's decoherence
must be a limiting factor on the number of operations that can be
performed on the TLSs, no matter how long their decoherence times
are. However, an important point to note here is that in the above
design with $N$ CBJJs, only one or two of them are used during any
gate operation. Therefore, the decoherence time of the entire
system is of the order of $T_{\rm d}^{\rm CBJJ}$ {\it rather than}
$T_{\rm d}^{\rm CBJJ}/N$, which one would normally obtain when
using $N$ CBJJs as phase qubits.

It is worth noting that there is some control, albeit very
limited, over the properties of the TLSs (e.g. the number of TLSs
in a CBJJ was drastically reduced by using a different barrier
material \cite{MS+S}). They will not be the only naturally formed
objects to allow quantum manipulation. A case in point is the
observation of Rabi oscillations in self-assembled quantum dots
\cite{Muller}.

{\it Summary.---\/} In conclusion, we have demonstrated that
quantum two-level systems (TLSs) naturally occurring in CBJJs can
be used as qubits. The one- and two-qubit gates, initialization
and readout can be readily performed, and the system can be scaled
beyond a single CBJJ. Being microscopis objects, TLSs have a
higher probability of possessing long decoherence times, which are
only weakly affected by the TLS-CBJJ interactions, due to that
interaction being switched off for most of the time. The
tunability of the CBJJs compensates for our currently limited
control over the parameters of the TLSs. In any case, both our
numerical simulations and especially the observation of quantum
beats between TLS and CBJJ \cite{Simmonds} show that the
experimental realization of our scheme is within the reach of
current experiments \cite{MS+S}.

\begin{acknowledgments}

We are grateful to M. Grajcar and Y. X. Liu for valuable comments.
This work was supported in part by the ARO, LPS, NSA and ARDA
under AFOSR contract number F49620-02-1-0334; and also supported
by the NSF grant No.~EIA-0130383.  A.Z. acknowledges partial
support by the NSERC Discovery Grants Program. S.A. was supported
by a fellowship from the JSPS.

\end{acknowledgments}


\begin{thebibliography}{99}
\bibitem{Temperature} Note that the temperature does not have
to be smaller than $E_J$ (with the proper unit conversion). It
only has to be smaller than the typical inter-level spacing
($\sim$ 100 mK).
\end{thebibliography}
\end{document}